\documentclass{emulateapj}
\usepackage{url}
\usepackage{rotating}

 \usepackage{graphics,graphicx}
 \usepackage{psfig}

\newcommand{\Msun}{\mbox{$M_{\odot}$}}

\begin{document}
\title{The Effects on Supernova Shock Breakout and {\it Swift} Light Curves Due to the Mass of the Hydrogen-Rich Envelope}

\author{Amanda J. Bayless\altaffilmark{1, 2},  Wesley Even\altaffilmark{3}, Lucille H. Frey\altaffilmark{3}, Chris L. Fryer\altaffilmark{3, 4, 5}, Peter W. A. Roming\altaffilmark{1, 2, 6}, Patrick A. Young\altaffilmark{7}}

\altaffiltext{1}{Southwest Research Institute, Department of Space Science, 6220 Culebra Rd, San Antonio, TX 78238, USA}
\altaffiltext{2}{University of Texas at San Antonio, San Antonio, TX 78249, USA}
\altaffiltext{3}{Los Alamos National Laboratory, Los Alamos, NM 87545, USA}
\altaffiltext{4}{Physics Department, University of Arizona, Tucson, AZ 85721, USA}
\altaffiltext{5}{Physics and Astronomy Department, University of New Mexico, Albuquerque, NM 87131, USA}
\altaffiltext{6}{Department of Astronomy \& Astrophysics, Penn State University, 525 Davey Lab, University Park, PA 16802, USA}
\altaffiltext{7}{School of Earth and Space Exploration, Arizona State University,  411 N Central Ave, Phoenix, AZ 85004, USA}

\begin{abstract}
Mass loss remains one of the primary uncertainties in stellar
evolution. In the most massive stars, mass loss dictates the
circumstellar medium and can significantly alter the fate of the star.
Mass loss is caused by a variety of wind mechanisms and also
through binary interactions.  Supernovae are excellent probes of this
mass loss, both the circumstellar material and the reduced mass of the
hydrogen-rich envelope.  In this paper, we focus on the effects of reducing
the hydrogen-envelope mass on the supernova light curve, studying both
the shock breakout and peak light curve emission for a wide variety of
mass loss scenarios.  Even though the trends of this mass loss will be
masked somewhat by variations caused by different progenitors,
explosion energies, and circumstellar media, these trends have
significant effects on the supernova light-curves that should be seen
in supernova surveys.  We conclude with a comparison of our results to
a few key observations.
\end{abstract}


\section{Introduction}
Core-collapse supernovae are classified by the evolution of
their emission and their spectral features (for a review, see \citealt{fil05}).
These features have been attributed to
characteristics of the progenitor star, where type Ib, Ic, and all
type II supernovae are believed to arise from massive stars. The
collapse of the stellar core in these stars releases the energy to power the
supernova.  The different types of supernovae: IIP, IIL, IIn, IIb, Ib, Ic
supernovae are believed to represent stars with different radii and
mass loss \citep{arn96,heg03}.  In this basic theoretical
picture, a type IIP supernova is believed to arise from the explosion
of a massive giant star, the extended envelope providing a plateau
phase in the emission as the photosphere sweeps through the extended
giant envelope.  As this envelope is removed, the star becomes more
compact and the plateau phase disappears.  As the hydrogen-rich envelope is
removed, the star exhibits helium-like features, producing a type IIb.
If the hydrogen-rich envelope is entirely removed, the supernova becomes a
pure Ib supernova.

The foundation of this simple theoretical picture of supernova
classification is based on our understanding of the evolution of
stellar radii and mass loss in massive stars.  As massive stars evolve
into a giant phase, the star is susceptible to a wide range of
instabilities: e.g. opacity driven instabilities ($\kappa-$mechanism)
in the hydrogen ionization zone \citep{fry06,pax13}, or
shell burning instabilities producing pulsations \citep{heg03,arn11}.  These instabilities can
significantly alter both the stellar radius and mass loss.  Stellar
radii in massive stars are notoriously difficult to observe.  For
massive stars, the mass loss from winds can set the photosphere beyond
the nominal edge of the star, making it difficult to use massive star
observations to observe the stellar radius.  Supernova observations
suffer from some of these same limitations, but they have the potential
to provide an independent determination of stellar radii.

Similarly, supernovae may help us probe the nature of mass loss in
massive stars.  The simple picture of line-driven winds for massive
star mass loss has gradually given way to a more chaotic picture where
outbursts from the giant envelope play a dominant role in the total
mass lost from systems.  In addition, evidence continues to grow
showing that most massive stars are in binaries, a sizable fraction in
interacting binaries that will shed mass due to binary interactions
\citep{kob07,koc12}.  Indeed, the
leading models for well-studied nearby supernovae SN 1987a and SN1993J
and supernova remnants, such as Cas A, both invoke binary interactions and mass loss
\citep{pod93,you06}.  One of the standard
mass-loss scenarios from binary interactions occurs when the star
expands in a giant phase.  If it envelops its companion, tidal forces
and friction will eject the hydrogen envelope by tapping orbital energy
(see \citealt{iva13} for a review).  This ``common envelope''
phase persists until either the entire hydrogen-rich envelope is ejected or
the companion merges with the helium core of the expanding star.

Whether the mass-loss is caused by steady winds, violent outbursts, or
binary interactions, we expect nature to produce, for every individual
progenitor star, a range of envelope masses and radii.  This project
is designed as a first step in studying the role of mass loss and
stellar radii in supernova light curves.  Here we study a single
supernova progenitor and explosion energy, varying the hydrogen mass
and stellar radius to determine their effect on the observed supernova
light curve.  In this manner, we isolate the effects of radius and mass to
predict trends.  These trends can be used to tie supernova observations
to uncertainties in stellar mass loss and radii.

 For this paper, we discuss the effects of removing hydrogen from the outer layers of the star on the evolution of SNe events.  We remove the hydrogen in two ways: 1) by removing mass in bulk from the outer shell, which also changes the radius, and 2) by altering the density of the outer shells. 
This simulates some of the many was in which a star can lose mass, e.g. winds, binary interactions, etc.  As a star loses hydrogen, it is expected that the SN would transition from a Type II/IIP to Type IIb and eventually to a Ib/c.  In this paper, we study the changes in light curves as this transition takes place.

\section{Supernova Models}

Supernova light curves depend upon a wide range of physical effects
and stellar characteristics including the structure of the stellar
core, the circumstellar medium, the explosion energy and the asymmetry
of the supernova.  In this project, we focus on the effects of mass
loss from the hydrogen-rich envelope on the supernova light curve.  As
such, our study will use a single progenitor with a standard $r^{-2}$
wind profile for the circumstellar medium.  This study will use a
single, spherically symmetric explosion.  Before we discuss our light
curve results, we will review our progenitor and supernova explosion
model.

\subsection{Progenitor}

\begin{figure*}
\center 
\includegraphics[scale=0.50]{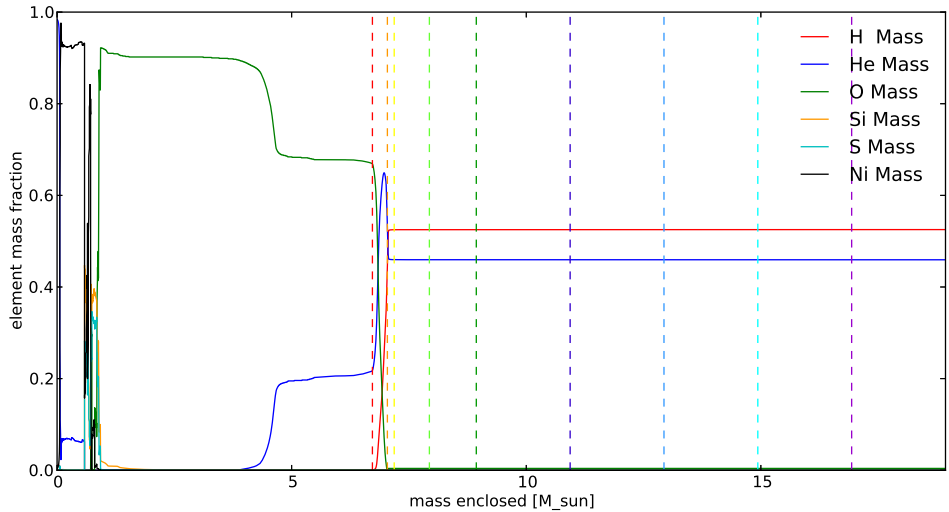}
\caption{Material mass fractions as a function of mass for our 23\Msun\ progenitor, post-explosion.  The dashed vertical lines corresponds to the outer layer of the star after shell mass removal for the models described in Table \ref{table:models}}
\label{fig:abundances}
\end{figure*}

The light curve models in this paper use a 23 \Msun\ progenitor
produced by the TYCHO \citep{you05} stellar evolution code.  This model
takes advantage of the newly revised mixing algorithm in the TYCHO
code.  TYCHO uses OPAL opacities \citep{igl96,ale94,rog96},
a combined OPAL and Timmes equation of state (HELMHOLTZ; \citealt{tim99,rog02}), gravitational settling, diffusion \citep{tho94},
general relativistic gravity, automatic rezoning, and an adaptable
nuclear reaction network with a sparse solver. A 177 element network
terminating at $^{74}$Ge is used throughout the evolution. The network
uses the latest REACLIB rates \citep{rau00, ang99,ili01,wie06}, weak
rates from \citet{lan00}, and screening from \citet{gra73}. Neutrino
cooling from plasma processes and the Urca process is included. Mass
loss uses updated versions of the prescriptions of \citet{kud89} for
OB mass loss, \citet{blo95} for red supergiant mass loss, and
\citet{lam02} for WR phases. It incorporates a description of
turbulent convection \citep{mea07, arn09, arn10, arn11} which is based
on three dimensional, well-resolved simulations of convection
sandwiched between stable layers, which were analyzed in detail using
a Reynolds decomposition into average and fluctuating quantities.  It
has no free convective parameters to adjust, unlike mixing-length
theory. The inclusion of these processes, which approximate the
integrated effect of dynamic stability criteria for convection,
entrainment at convective boundaries, and wave-driven mixing, results
in significantly larger extents of regions processed by nuclear
burning stages.

This new mixing algorithm can significantly alter the helium shell.  
Figure~\ref{fig:abundances} shows the distribution (mass fractions) 
of key elements in the post-explosion star.  Notice the peculiar 
helium shell abundances.  The mixing algorithm in TYCHO produced 
extensive helium burning, converting most of the helium in this shell 
to oxygen.  When the hydrogen-rich envelope is removed exposing the He-shell, this object will look 
much more like a type Ic supernova \citep{fre13b}.

\subsection{Supernova Explosion}
The core collapse itself was done with a one-dimensional Lagrangian code developed by \citet{her94}.  This code includes three-flavor neutrino transport using a flux-limited diffusion calculation and a coupled set of equations of state and nuclear networks to model the
wide range of densities in the collapse phase (see \citealt{her94, fry99} for details).  After collapse and bounce, the proto-neutron
star is removed (and replaced with a hard boundary).  We continue the
evolution by injecting energy at the surface of the proto-neutron star to
drive an explosion.  The duration and power of this energy injection
can be modified to mimic rapid and delayed explosions \citep{fry12}.
For this paper, we produced a rapid explosion with a higher than average energy, $E=5\times10^{51}$~erg, in a 
spherical explosion.  

We follow this explosion until the shock is at $2.9 \times 10^{10}$~cm.  As we expect
for many stars above 20~\Msun\ much of the inner material is not given
enough energy to escape the gravitational pull of the proto-neutron
star and it falls back onto the neutron star.  We model the fallback
by allowing it to pile up on the proto-neutron star surface and cool
via neutrino emission.  When the fallback material density exceeds
$10^9$~g~cm$^{-3}$, we assume neutrino emission is able to cool the
material quickly (on the timescale of our hydrodynamics timestep) and
assimilate it into the proto-neutron star.  

The explosion code includes a small 17-isotope network.  But we 
post-process all the ejected material with the 489-isotope TORCH\footnote{http://cococubed.asu.edu}
code.  Note in Figure \ref{fig:abundances} that there is  $7.3\times10^{-3}$ \Msun\ of $^{56}$Ni in the core 
because most of the inner material fell back and was accreted onto 
the proto-neutron star.  The low $^{56}$Ni yield alters the light-curve, 
especially past peak and must be considered when comparing our light-curve 
calculations to observations.    To demonstrate the importance of the 
$^{56}$Ni yield, we include two models with enhanced $^{56}$Ni in Section 5.

\subsection{Hydrogen-Rich envelope and Mass Loss}

\begin{figure*}
\center 
\includegraphics[scale=0.45]{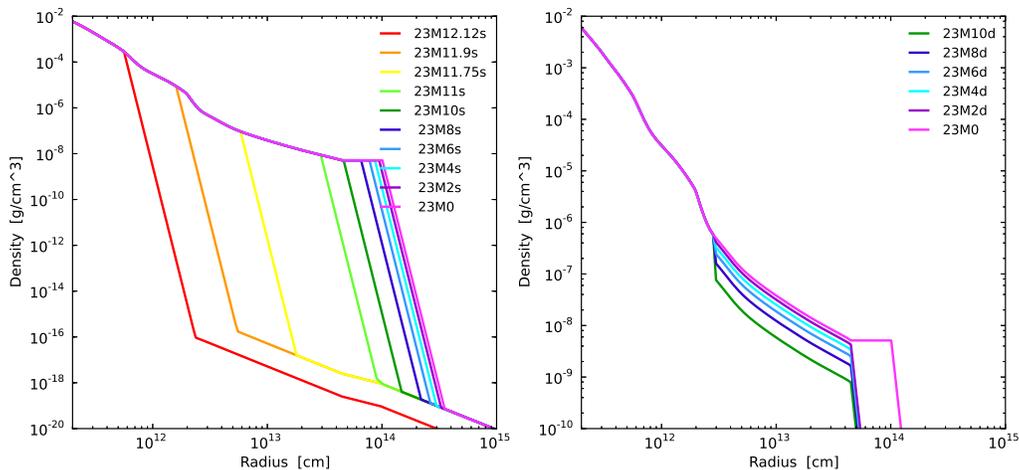}
\caption{
 The initial density profiles for the simulations presented in this paper.  The figure on the left displays the simulations where the mass was removed by removing layers from the star (shell models) and the figure on the right displays simulations where the density in the outer layer was decreased to remove additional mass after an initial shell was removed(density models).}
\label{den}
\end{figure*}

The 23~\Msun\ solar metallicity star is evolved up to the onset
of core collapse in TYCHO.  The single, non-rotating 23~\Msun\ star
evolves normally as a red supergiant.  This final star mass
is 18.9~\Msun, with a 11.9~\Msun\ H-rich envelope, containing 6.2~\Msun\ of hydrogen and 5.6~\Msun\ of helium.  This progenitor has an unusually high helium composition containing 52\% hydrogen and 47\% helium. Mass loss remains
one of the primary uncertainties in our understanding of the hydrogen
envelope.  The focus of this paper is to study the effects of enhanced 
mass loss.

For this study, we use the exploded TYCHO simulation as a base and then remove
mass using two different procedures: 1) a shell mass removal where the
outer layers of the star are removed from our model, reducing both the
H shell mass and the stellar radius, and 2) a density-decrease mass removal
where we reduce the density of the H shell, reducing the mass, but not
the radius of the star.  Mass loss occurs through a variety of
mechanisms: winds, stellar instabilities, binary interactions.
Although one can imagine scenarios where mass is removed and the star
shrinks or expands significantly, these two procedures span a
relatively wide range of the possible outcomes from mass loss,
providing a basic idea of the trends from mass loss.
Table \ref{table:models} shows the range of models used in this study and Figure \ref{den} displays the initial density profiles for all models.

\begin{table}[htb]
\footnotesize
\center
\caption{Summary of Explosion Models}
\label{table:models}
\begin{tabular}{lccc}
\hline
Model\tablenotemark{c} & Radius & Shell Mass & Removal \\
Name & (cm) & Removed [\Msun] & Mechanism \\
\hline
\hline
23M0       & $1.01\times10^{14}$ & 0.0   & Baseline \\
23M2s      & $9.45\times10^{13}$ & 2.0   & Shell\tablenotemark{a} \\
23M4s      & $8.64\times10^{13}$ & 4.0   & Shell\tablenotemark{a} \\
23M6s      & $7.78\times10^{13}$ & 6.0   & Shell\tablenotemark{a} \\
23M8s      & $6.58\times10^{13}$ & 8.0   & Shell\tablenotemark{a} \\
23M10s     & $4.63\times10^{13}$ & 10.0  & Shell\tablenotemark{a} \\
23M11s     & $2.93\times10^{13}$ & 11.0  & Shell\tablenotemark{a} \\
23M11.75s  & $5.56\times10^{12}$ & 11.75 & Shell\tablenotemark{a} \\
23M11.9s   & $1.61\times10^{11}$ & 11.9  & Shell\tablenotemark{a} \\
23M12.12s  & $5.59\times10^{11}$ & 12.12 & Shell\tablenotemark{a} \\
23M10.42d  & $4.49\times10^{13}$ & 10.42 & Density\tablenotemark{b} \\
23M10.72d  & $4.49\times10^{13}$ & 10.72 & Density\tablenotemark{b} \\
23M11.01d  & $4.49\times10^{13}$ & 11.01 & Density\tablenotemark{b} \\
23M11.31d  & $4.49\times10^{13}$ & 11.31 & Density\tablenotemark{b} \\
23M11.60d  & $4.49\times10^{13}$ & 11.6  & Density\tablenotemark{b} \\
\hline
\end{tabular}

\tablenotetext{1}{For our Shell Models, we remove mass starting from outer layers of the star, both removing mass and shrinking the stellar radius.}
\tablenotetext{2}{For our density models, we lower the density in the hydrogen shell, removing H-shell mass, but retaining the stellar radius.}
\tablenotetext{3}{All Models use a 23 \Msun\ star with $5\times10^{51}$\, erg explosion and 0.0073 Ni mass.}

\end{table}

\subsection{Light Curve Code}
Before the shock has reached the helium shell, the explosion is mapped
into RAGE (Radiative Adaptive Grid Eulerian; \citealt{git08}) to
model the coupling between matter and radiation as the shock breaks
out of the star and interacts with the interstellar medium.  The density structure
and abundance of the interstellar medium can greatly affect the resulting  
simulated light curves, even for low masses of interstellar material.  In order to minimize
the impact of the circumstellar material on the light curves a low density wind resulting 
from a mass loss rate of $10^{-7}$~\Msun/yr and velocity of  $10^8$~cm/s was
implemented in all models.  Even at this extremely low mass loss rate the wind will still alter the light curves at breakout, but this allows for a nearly single parameter numerical study over the time frame of which Swift data is present.  \citet{bal11} demonstrate that the radiation energy in the shock at breakout is approximately equal to the mass loss rate to the $\frac{16}{21}$ power.  Therefore, increasing the mass loss rate to $10^{-5}$~\Msun/yr would result in a $\sim 40$ times brighter breakout in models where the breakout occurs in the wind, which would mask some of the differences we are studying here.

Although the RAGE code is capable of simulating the explosion in 1-, 2-, and
3-dimensions with multi-group flux-limited diffusion, for these
calculations we model the explosion in 1-dimension using a gray
opacity scheme.  These calculations include the shock heating effects
on the temperature and the radiation effects on the hydrodynamics that
are critical in modeling shock break-out and the peak supernova
light-curve for most core-collapse supernovae.  To produce the
detailed spectra needed to produce light-curves in different bands, we
post-process these radiation-hydrodynamics calculation using the
opacities from the SESAME database.  The post-process technique,
developed by \citet{fry09,fry10} is described in detail
in \citet{fre13}.

\section{Shock Breakout}
\begin{figure}
\center
\includegraphics[scale=0.45]{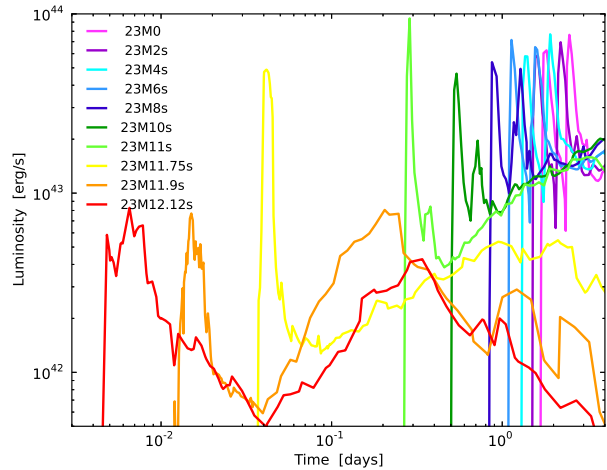}
\caption{Bolometric light curves for the shell removal models at shock breakout. } 
\label{fig:bolc}
\end{figure}

\begin{figure*}
\center
\includegraphics[scale=0.55]{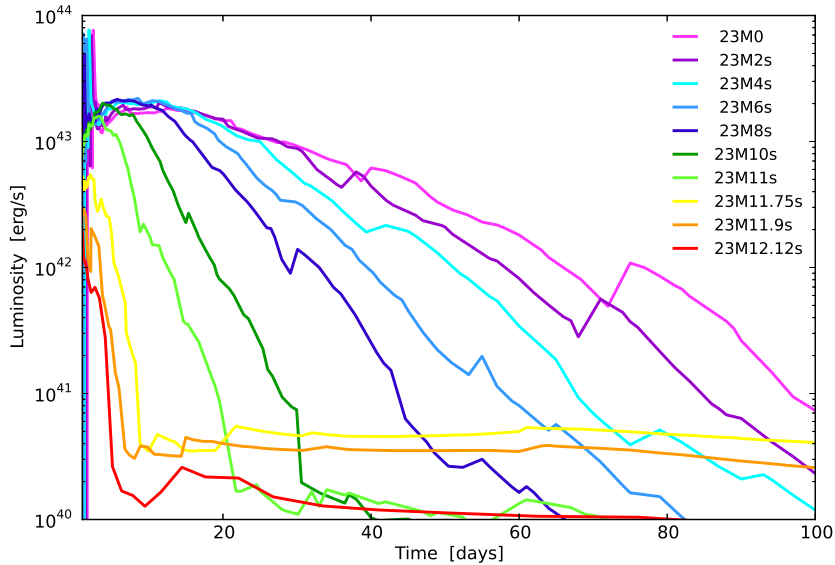}
\caption{Bolometric light curves for the shell removed models showing the transition as more mass is removed.}
\label{f4}
\end{figure*}

The supernova explosion launches a shock from the core that travels outward through the star.  When this shock reaches the surface it rapidly accelerates and increases in temperature as it traverses the 
steep density gradient between the star and the surrounding circumstellar material (see Figure \ref{den}).  Prior to shock breakout the radiation is trapped within the hydrodynamic shock because the diffusion time 
for the photons is much greater than the hydrodynamic time scales.  If there is little circumstellar material, at shock breakout the photon diffusion timescale decreases below the hydrodynamic timescale and 
results in a short but powerful ($\gtrsim 10^{44} \mathrm{erg/s}$) burst of photons escaping from the outer edge of the shock.  The majority of this energy will be released in the hard UV, X-rays, and gamma rays.
In stars with dense circumstellar material the photon breakout can be delayed significantly from the shock breakout at the surface of the star \citep{col74,dop87,ens92,tan01}.
Figure \ref{fig:bolc} shows bolometric light curves for shock breakout of all 10 shell removal models.  The breakout peaks last from $< 10^{-2}$ s for 23M12.12s
to $\sim1$ day for 23M0.  While brief, the peak bolometric luminosity reached during shock breakout can be over an order of magnitude greater than what is typically labeled as
peak in a standard SN light curve.  The early time low amplitude modulations in this figure and in the following figures are created by multiple small shocks.  These are likely an
artifact of the 1D simulations and in multi-dimensions turbulence within the ejecta will cause dissipation in these shocks.  The decrease in breakout time with increasing mass removal results mainly from the decrease in radius, although the relation is not linear
due to the deceleration of the shock as it sweeps up mass while propagating through outer layers.  Figure \ref{f4} shows the transition as mass is removed from the shell.  

Shock breakout luminosity is commonly estimated by applying the Stefan-Boltzmann equation at the point when the shock first reaches the $\tau=1$ radius, resulting
in luminosity being a simple function of radius and shock temperature \citep{svi12}.  This simple back of the envelope calculation can deviate from the actual breakout luminosity
by orders of magnitude because there is no single radius where radiation breakout occurs.  Instead each photon energy has a unique opacity resulting in a potentially wide range  in $\tau_\nu=1$ radii and temperatures.

\begin{figure*}
\center
\includegraphics[scale=0.45]{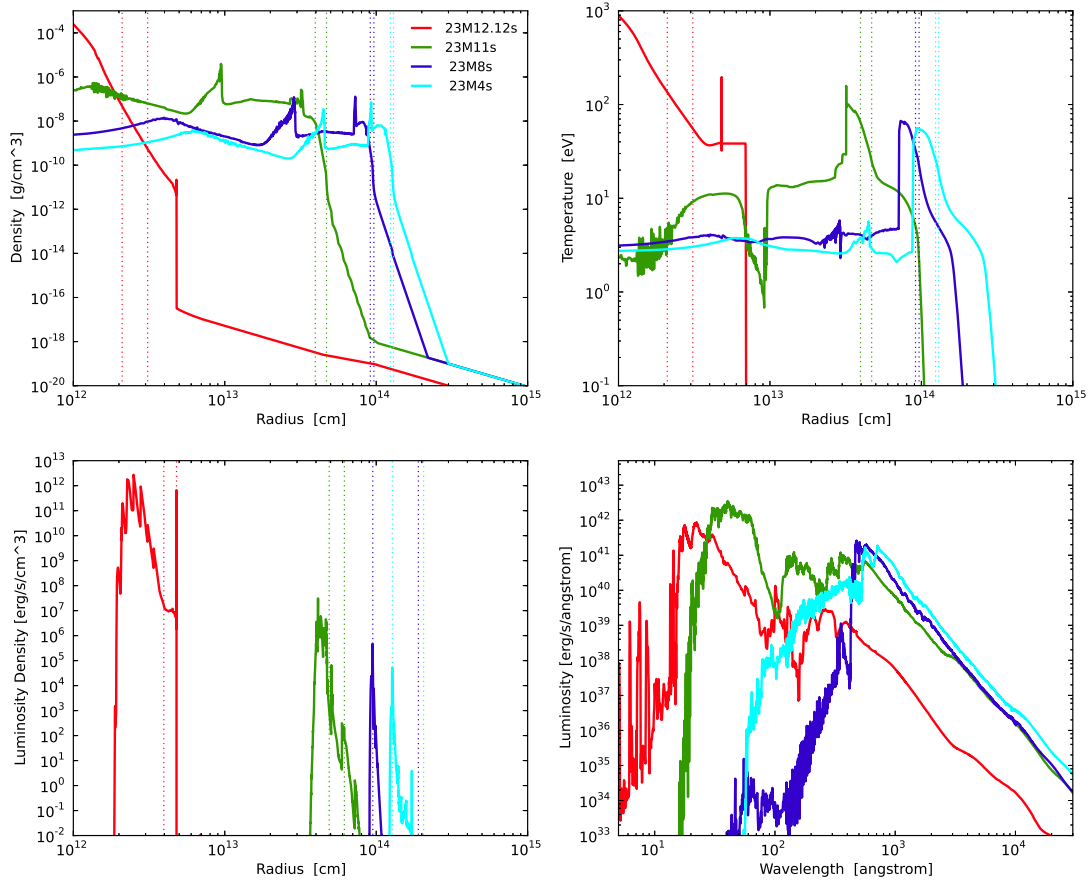}
\caption{The density (top left), temperature (top right), luminosity density (bottom left), and spectra (bottom right) for four different simulations at the time
that corresponds to peak breakout luminosity.   The vertical dashed lines in the top plots enclose the volume that is the source of 99\% of the
luminosity.  The vertical dashed lines in the bottom left plot represent $\tau=1$ surfaces at two wavelengths near the peak of the spectra for that model 
(10\AA\ \& 20\AA\ for 23M12.12s, 20\AA\ \& 40\AA\ for 23M11s, 100\AA\ \& 1000\AA\ for 23M8s \& 23M4s). }
\label{fig:bo}
\end{figure*}

This is illustrated in Figure \ref{fig:bo}, where we show the $\tau=1$ surface calculated at two different wavelengths for each model at the time of peak breakout luminosity. Figure \ref{fig:bo} also shows that temperature varies by a factor of a few within the emitting region, with the temperature increasing with smaller radii.  This
gradient in temperature is also a factor in creating a large emitting region as more luminosity is created in the higher temperature region and therefore can contribute to the observed 
luminosity even though it is at a higher $\tau$.  The spectra in Figure \ref{fig:bo} reflect the higher breakout temperature with the most compact explosion peaking in the X-ray, while the more extended stars have spectral energy distribution that peak in the UV.

\begin{table*}[t]
  \centering
  \caption{Breakout Parameters}
  \begin{tabular}{lccccc}
  \hline
  model & $R_{bo}$ $[\mathrm{cm}]$ & $T_{bo}$ $[\mathrm{eV}]$ & $L_{an}$ $[\mathrm{erg}/\mathrm{s}]$ &  $L_{sim}$ $[\mathrm{erg}/\mathrm{s}]$ & $\rho_{column}$ $[\mathrm{g}/\mathrm{cm}^2]$ \\
  \hline
  \hline
  23M12.12s     &  $2.96 \times 10^{12}$  &  36.7   &   $3.7 \times 10^{44}$    &  $8.3 \times 10^{42}$    &  $1.7 \times 10^{ 1} $ \\
  23M11s        &  $6.16 \times 10^{13}$  &  13.3   &   $1.6 \times 10^{45}$    &  $9.4 \times 10^{43} $   &  $1.4 \times 10^{-2} $ \\
  23M8s         &  $1.42 \times 10^{14}$  &  3.89   &   $6.0 \times 10^{43}$    &  $5.4 \times 10^{43} $   &  $8.6 \times 10^{-3}$ \\ 
  23M4s		&  $1.88 \times 10^{14}$  &  5.11   &   $3.1 \times 10^{44}$    &  $3.9 \times 10^{43} $   &  $5.5 \times 10^{-4}$ \\ 
  \hline
  \end{tabular}
    \label{bot}
\end{table*}
\vspace{0.1in}

In Table \ref{bot} we compare the single radius black-body approximation to the simulated results shown in Figure \ref{fig:bo}.  A luminosity-weighted averaged 
radius and temperature are calculated from the models in Figure \ref{fig:bo}.  The analytic luminosity is constructed by extracting the radius and temperature at 
the location of the $\tau=1$ surface at 20\AA\ from the simulation.  This radius and temperature pair are used to calculate the luminosity for an ideal 
black-body sphere.
The difference between the analytic and the simulated bolometric luminosities are the most extreme for the more compact stars, where the discrepancy is over two orders of magnitude.  The 23M12.12s and 23M11s are emitting light from an average depth such that there is $\sim 2 \times 10^1$ of column density that the photons must travel through to escape.  Only photon energies with opacities $\lesssim0.1$ g/cm$^2$ will be able to escape from this region causing non-blackbody emission, as can be seen in the spectra in the bottom right panel of Figure \ref{fig:bo}. In contrast, the more massive models only have a column density of $\sim 10^{-3}$ g/cm$^2$ from the emitting region resulting in much less extinction and a more reliable black-body fit.

The wavelength dependence of the opacity is also responsible for the transition from the single peaked breakout light curve for the most stripped stars to the double peaked breakout in the more massive stars.  The double peak results from higher energy photons with lower opacity escaping from deeper within the surface of the star than the lower energy photons that can only begin to free stream closer to the surface.

\begin{figure*}
\center
\includegraphics[scale=0.55]{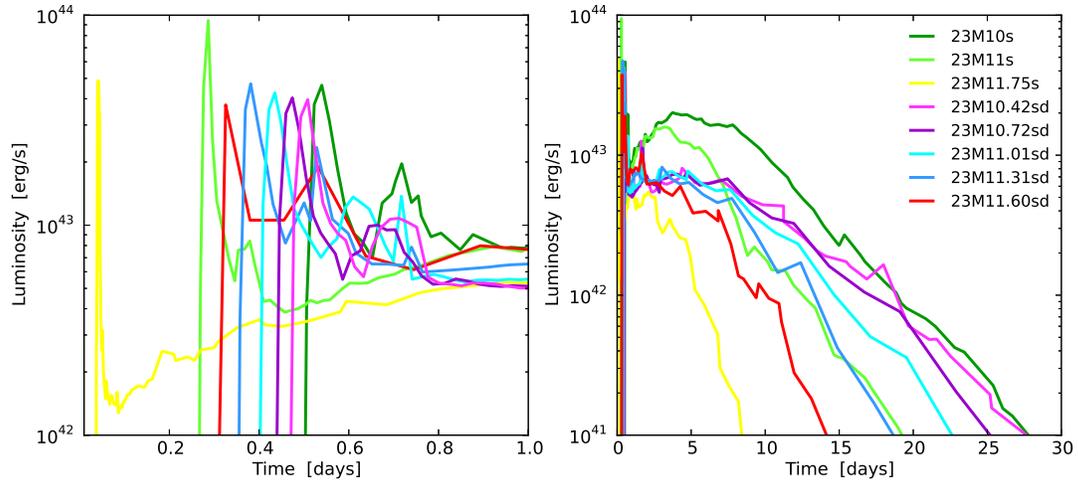}
\caption{Comparison of bolometric light curves between the shell removed and density altered models. }
\label{bod}
\end{figure*}

The left panel of Figure \ref{bod} compares the breakout light curves for the mass and shell removal models.
In simulations, the density altered models, which have a larger radii, are dimmer than the corresponding shell removal models at peak brightness, but have a slower decay.
The difference in increased peak brightness is largest for models with the least amount of mass removed, which have the most material for the shock to interact with as 
it breaks out of the surface of the star.

\begin{figure*}
\center
\includegraphics[scale=0.40]{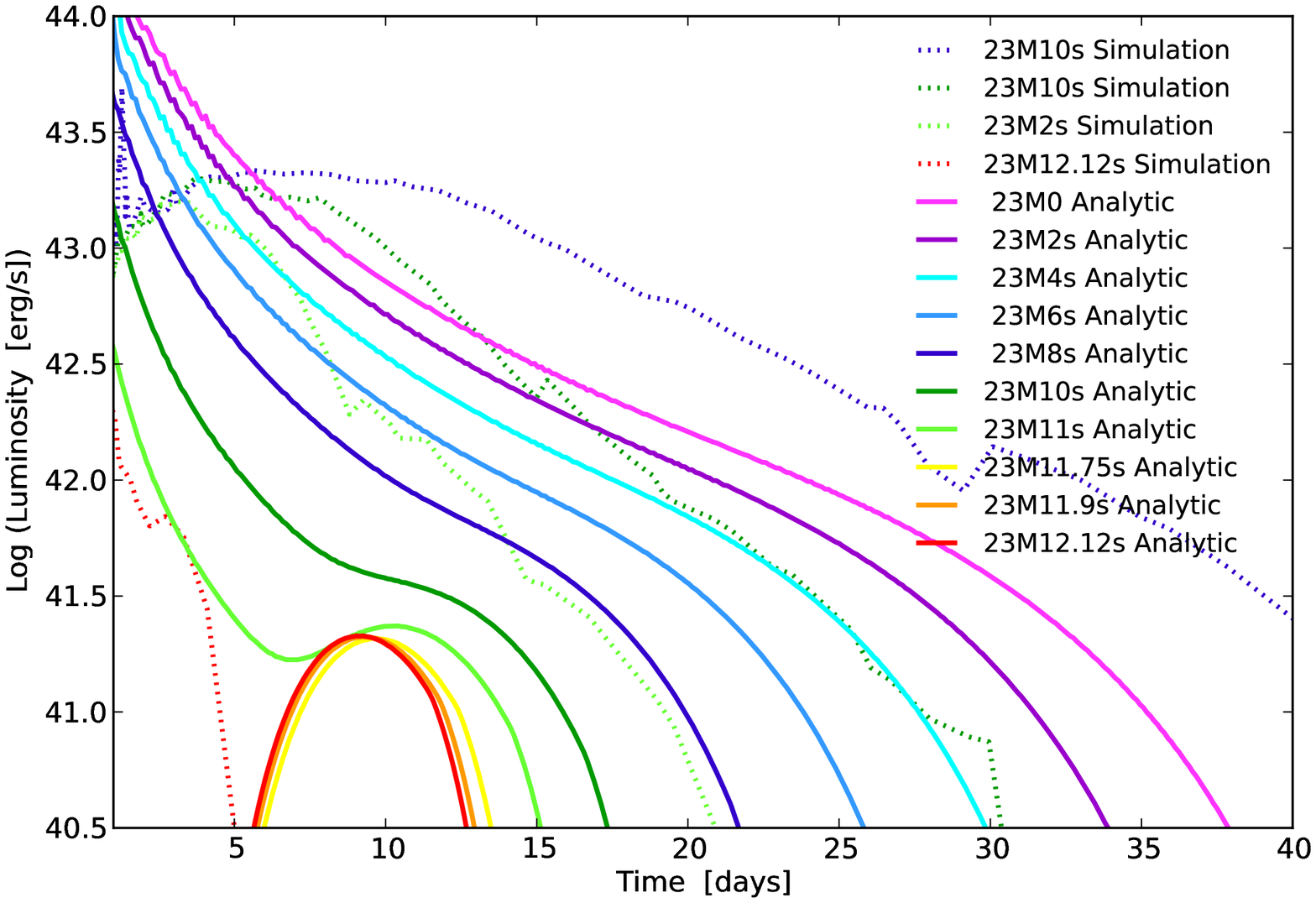}
\caption{A comparison of light curves generated from rage simulations (dotted) with a simple homologous expansion plus diffusion model}
\label{analytic}
\end{figure*}

\section{Light Curves}
We present multiple simulated light curves from the models described above. Bolometric light curves are shown in Figure \ref{fig:bolc}.
The models with 0 to 4 \Msun\ removed are very similar for the first 20 days and the early time light curves do not start to deviate until 6 \Msun\ of material is removed.
Since all models have an
identical explosion, the differences observed in the light curve are solely a result of interactions with the outer layers of the star, which for 23M0 to 23M4s are
not dramatic with only a 14\% change in radius and 21\% change in mass between the 3 models.  However, when $\ge$ 6 \Msun\ is
 removed the slope of the fall-off changes dramatically and when $\gtrsim$ 11 \Msun\ the peak also begins to rapidly drop with mass removal.  
To better understand the importance of shock heating, we have
developed a simplified model of the light-curve evolution based on the
work of \citep{arn80,arn82}.  An analytic method can be developed by
making a number of simplifying assumptions including an expansion that
is homologous ($v \propto r$), the radiation energy dominates the
energy equation (the total energy $E=aT^4V$ where $a$ is the radiation
constant, $T$ is the temperature, and $V$ is the volume), single-group
gray opacity, and the diffusion equation is adequate model the
light-curves.  This approach\citep{arn80,arn82} has been used
successfully to build intuition and make a first pass at the light
curves from explosions.  Our ``test'' code numerically evaluates the
light-curve with these same assumptions: we assume homologous
expansion of the supernova after the shock breaks out of the star, we
include energy deposition from the decay of $^{56}$Ni (high
$^{56}$Ni model), and diffusion scheme using a constant gray opacity to
transport energy.  The initial conditions are set by approximating the
initial conditions in our full calculations: we assume a constant
density profile and a constant temperature of 30eV.  Early emission
and adiabatic cooling quickly deplete this initial energy during shock
breakout and the late-time light-curve is entirely powered by the
decay of $^{56}Ni$ and its daughter products.  This code matches the
analytic approximations, but can easily be modified to include
additional physics to compare to our more detailed light-curve code.

These light-curve calculations are plotted in Figure \ref{analytic}.  The gray opacity and the assumption of a single photosphere produces a very sharp breakout signal that is both higher and shorter in duration than our simulations.  A major assumption in these analytic calculations (and in many radiation-transport only calculations) is that $^{56}$Ni is the dominant energy source for the light curve.  However, shock heating can dominate the energy source in many core-collapse explosions.  Once we launch our explosion, we assume no further shock heating (either from wind or reverse-shock interactions).  Comparing with our full simulations, we see that shock heating plays a dominant role in the light curve, especially at late times.  Additionally, the light curves from a full simulation tend to be better fits with a simple model that had less mass stripped than its direct comparison model.
 
\begin{figure}[ht]
\center 
\includegraphics[scale=0.40]{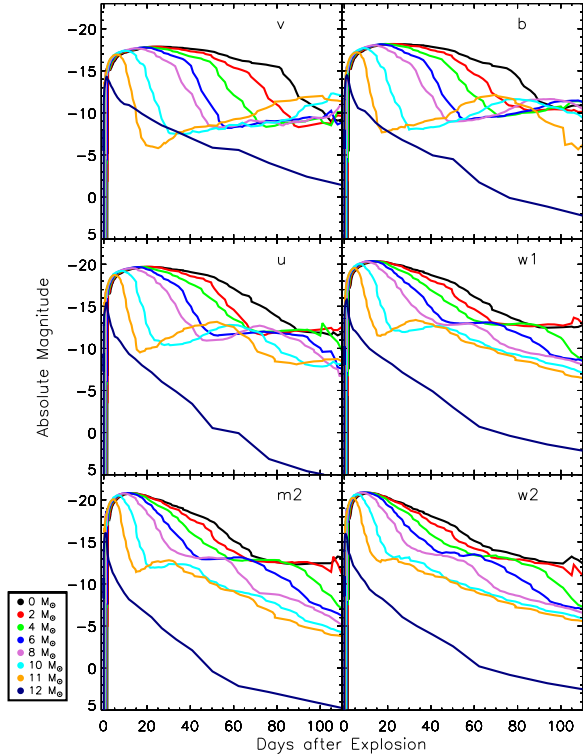}
\caption{The light curves for the models in Table \ref{table:models}.  The six band passes are the UVOT filters.  The general trend has the UV light curves of the larger star (less removed) peaking later and having a faster fade out.  The optical light curves in all the models are similar within the first month.}
\label{modelLC2}
\end{figure}

\begin{figure}[ht]
\center 
\includegraphics[scale=0.40]{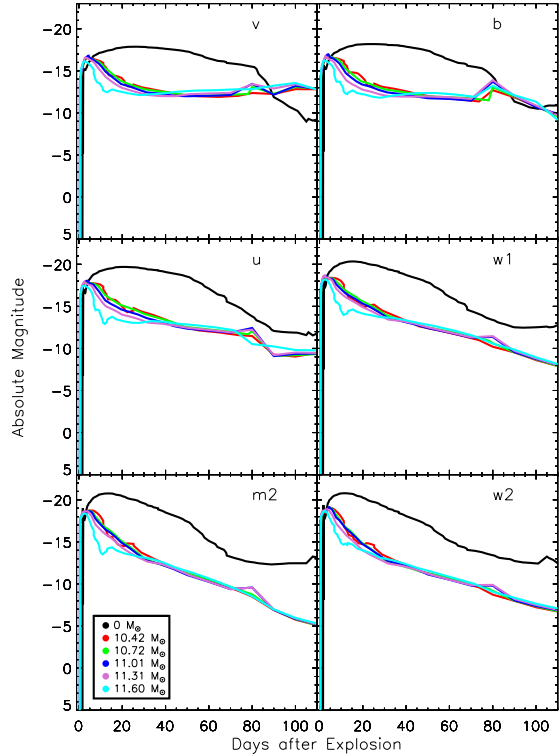}
\caption{The light curves for the altered density models in Table \ref{table:models} in the same UVOT filters as in Figure \ref{modelLC2}.  These show the same general trend, but have a slower fall off post-peak.}
\label{modelDEN}
\end{figure}

Our simulations are also shown in the Swift \citep{geh04} UVOT \citep{rom00,rom04,rom05}
band passes. Figure \ref{modelLC2} shows the shell model light curves and Figure \ref{modelDEN} shows the altered density models.
The central wavelengths for the {\it Swift} filters are {\it uvw2} ($\lambda_c =1928$ \AA), {\it uvm2} 
($\lambda_c =2246$ \AA), {\it uvw1} ($\lambda_c =2600$ \AA), {\it u} ($\lambda_c =3465$ \AA), {\it b} 
($\lambda_c =4392$ \AA), and {\it v } ($\lambda_c =5468$ \AA; \citealt{bre10}).
We note that the {\it Swift} $u$-band is not a traditional U-band, but has a bluer response than the atmospheric cut-off. 
The models with $<$ 4 \Msun\ removed show a very similar
rise and fall in the UV and a gradual rise and fall in the optical, as typical for a Type IIP SNe \citep{bro09}. 
As more mass is removed, the slope of the post-peak fall off becomes steeper and begins to resemble Type IIb and Type Ib/c SNe. For the case of 12 solar masses removed in Figure \ref{modelLC2}, the entire hydrogen shell has been removed. The time of peak luminosity and 
the slope of the post-peak decline for the baseline model and models 23M6s and 2311.01d are summarized in Table \ref{tablc}. 
These models are shown because they are representative of the slope changes as mass is lost. 

\begin{figure*}[ht]
\center 
\includegraphics[scale=0.45,angle=0]{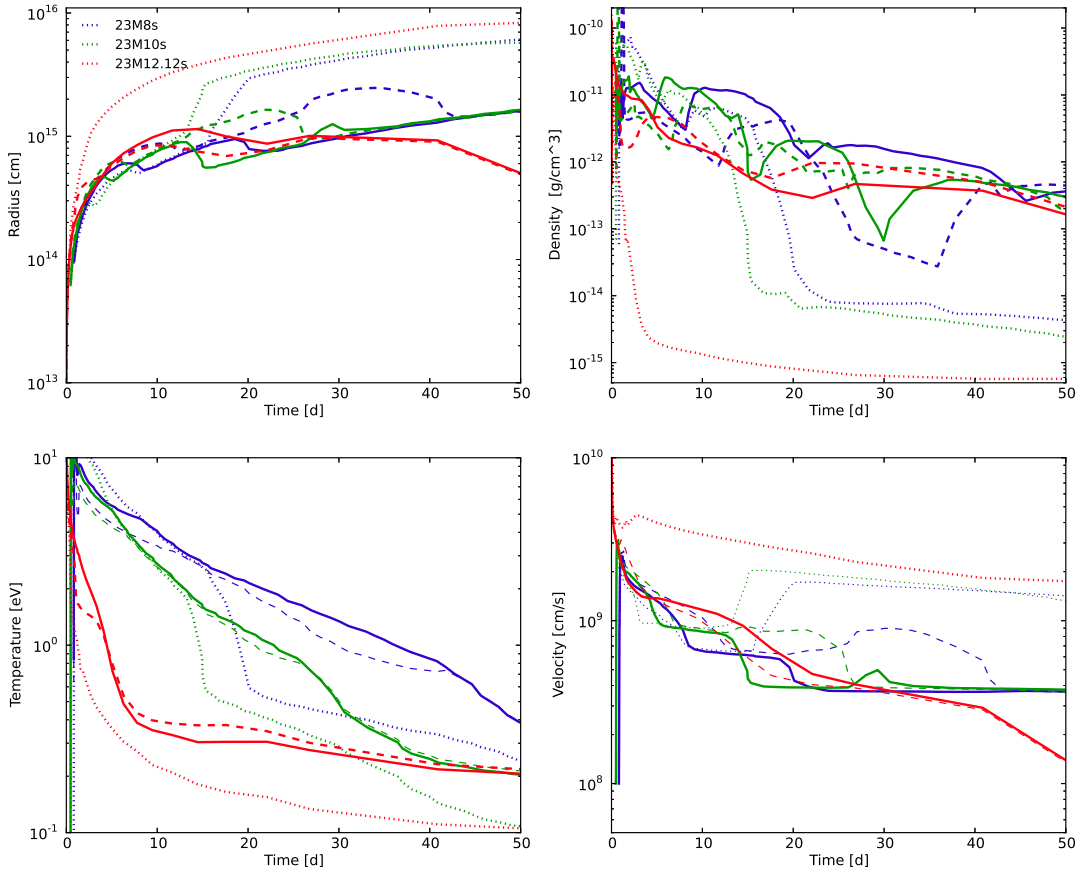}
\caption{Time-Series evolution of the radius, density, temperature, and velocity. The time series evolution of the position, density, temperature, and velocity of the $\tau = 1$ surface for 3 different {\it Swift} bands for 3 different models.  The dotted lines correspond to the {\it uvm2}-band, the solid lines to the {\it u}-band, and the dashed lines to the {\it v}-band.}  
\label{tau1}
\end{figure*}

 Figure \ref{tau1} shows how the model's radius, temperature, density, and velocity change with time.  The upper left panel of Figure \ref{tau1} displays the position of the $\tau = 1$ surface at 3 different Swift wavelengths ($uvm2$, $u$, and $v$) for the 23M8s, 23M10s, and 23M12.12s models.  The upper right, lower left, and lower right panels present the density, temperature, and velocity in the simulation corresponding to the $\tau=1$ surface.    From 0 to 10 days the $\tau = 1$ surface is expanding and rapidly cooling resulting in a rapid decline in the light curve.  At $\sim$10 days the $\tau=1$ surface drops deeper into the ejecta and  reveals higher temperature material causing a rise in the light curve.  In the $uvw2$-band there is a similar, but less dramatic rise in the light curve and a correspondingly smaller rise in the temperature at the $\tau=1$ surface.  It is also noted that the $\tau=1$ surface for the UV bands and the optical bands represent very different velocities and will create different Doppler shifts for lines across these wavelengths.  

\begin{table*}[htb]
\footnotesize
\center
\caption{Peak \& Slope of Selected Light Curve Models}
\label{tablc}
\begin{tabular}{c|ccc|ccc|ccc|}
\hline
 Mass Removed =  &\multicolumn{3}{c|}{0 \Msun}	& \multicolumn{3}{c|}{6 \Msun, shell } & \multicolumn{3}{c|}{11.01 \Msun, density }\\ 
\hline
Band Pass 	&	Peak\tablenotemark{a} & Slope 1\tablenotemark{b}  & Slope 2\tablenotemark{c}  & Peak & Slope 1 & Slope 2 & Peak & Slope 1 & Slope 2\\	
\hline
\hline
$uvw2$ 		& 	12	&	0.11	& 0.08	&	9 	&	0.24	& 0.11 	&	 6	&	0.21	& 0.07	 \\
$uvm2$ 		&	14	&	0.09	& 0.09	&	10	&	0.26	& 0.11	&	 7	&	0.17	& 0.10	 \\
$uvw1$ 		&	16	&	0.07	& 0.10	&	14	&	0.21	& 0.08	&	 3 	&	0.17	& 0.08	 \\
$u$ 		&	21	&	0.04	& 0.12	&	16	&	0.14	& 0.05	&	 5	&	0.19	& 0.08	 \\
$b$ 		&	22	&	0.02	& 0.12	&	17	&	0.11	& 0.00	&	 5	&	0.21	& 0.04	 \\
$v$			&	26	&	0.02	& 0.14	&	20	&	0.10	& 0.00	&	 5	&	0.20	& 0.02	 \\
\hline  
\end{tabular}
\tablenotetext{1}{Peak is number of days past explosion.}
\tablenotetext{2}{Slope 1 is the average decline between the peak and about 40 days past explosion in mag/day.}
\tablenotetext{3}{Slope 2 is the average decline after about 40 days past explosion in mag/day.}
\end{table*}

\subsection{Comparisons to Normal Supernovae}

\begin{figure*}[ht]
\center 
\includegraphics[scale=0.55,angle=90]{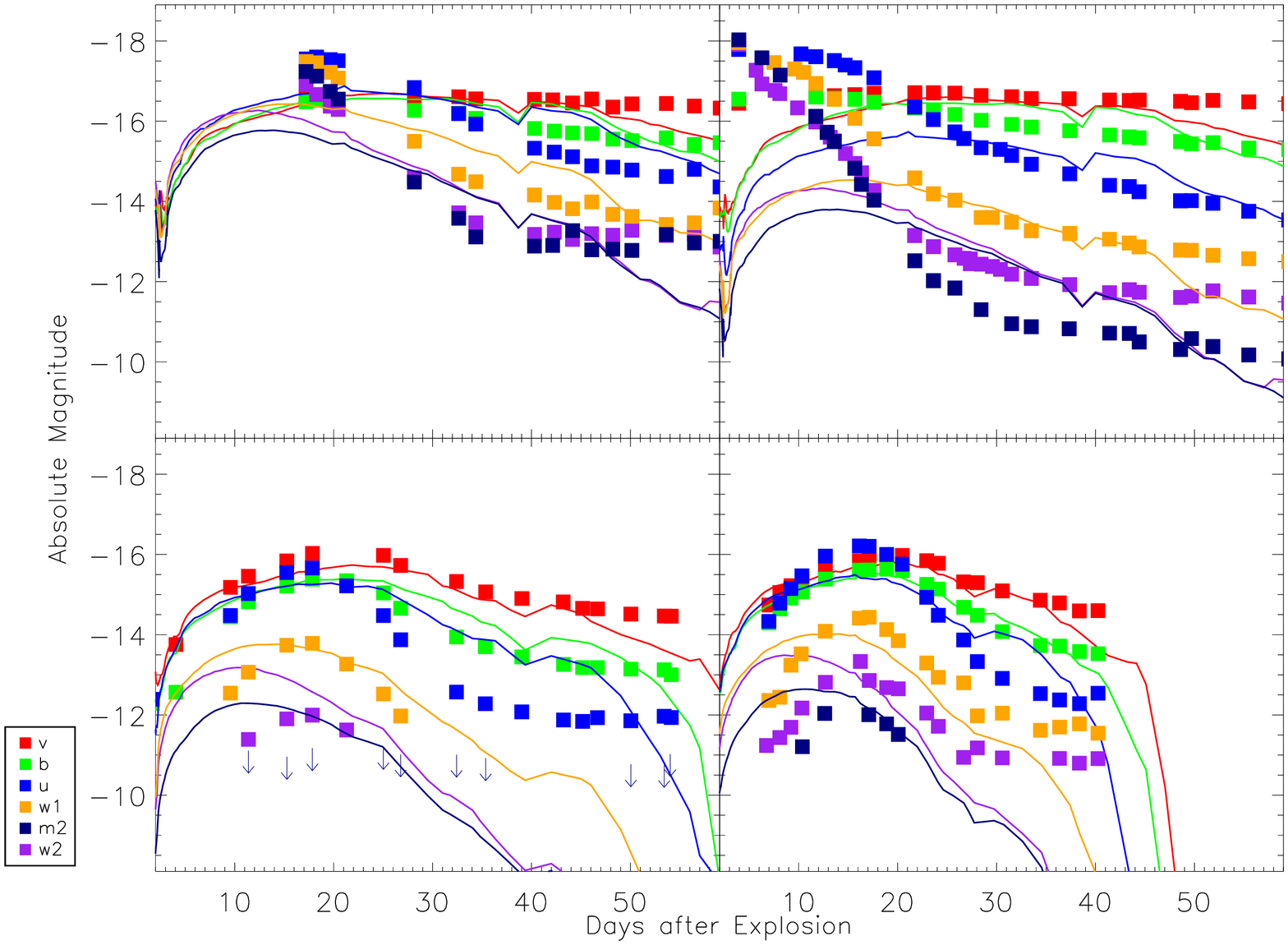}
\caption{Comparison of mass loss models to {\it Swift} observed SNe of several types. See text for details. {\it Top Left}-- The baseline model compared to Type IIP SN 2010F.  {\it Top Right}--The baseline model compared to Type IIP SN 2012aw. {\it Bottom Left}-- Model 23M4s compared to Type IIb SN 2008ax.  {\it Bottom Right}-- Model 23M6s compared to Type Ib SN 2007Y.}
\label{lcmodplot}
\end{figure*}


Figure \ref{lcmodplot} shows a sample comparison to {\it Swift} observations (see \citealt{pri13}). The error bars on the observations are approximately the size of the data points. Note that these models are not intended to fit these specific light curves, but are shown only to illustrate that we do match observed trends. The {\it Swift} data has been galaxy subtracted and a reddening law has been applied to the models based on \citet{car89} with $R_v=3.1$.  In general, the baseline model can be fit to a Type IIP and the model where hydrogen is removed can be fit to Type IIb and Type Ib.    However, there are still a few issues that require further modeling.  The overall model light curves are still too luminous after reddening, particularly in the UV and required adjustment to be comparable to the {\it Swift} observations.  This adjustment was to artificially add reddening by increasing $E(B-V)$ until the $v$ and $b$-bands were close to fitting and then to add an additional constant magnitude term to the UV and $u$-bands.  This needed to be done because the models produced 5 foe of energy, much more than an typical explosion.  Future models will have lower energies and be more realistic for the comparison to observations.

The top left panel shows the baseline model with the {\it Swift} light curves for the Type IIP SN 2010F \citep{maz10}.  SN 2010F does not have well constrained parameters allowing some adjustment in reddening and time of explosion.  The discovery of SN 2010F was on JD 2455209.8 with an earliest possible date of explosion on 2455189.5.  The galactic reddening is 0.095, but the host reddening is unknown \citep{maz10}. We are able to come close to fitting the post-peak slope until about day 40, but this required shifting the explosion date to an unrealistic 14.7 days {\it after} discovery, to JD2455224.5.  We also needed an additional reddening of 0.3 and required an additional constant added to the UV and $u$-band models of 1.3 mag and 1.0 mag respectively.  
The top right panel shows the same baseline model with the Type IIP SN2012aw \citep{bay13}. SN 2012aw was some what unique observationally in that its close proximity (10 Mpc, in M95) allowed lengthy UV observations.  These extended UV observations showed a flattening at late times, $>$ 30 days past explosion.   The baseline model is matching the general late time slope observed and the {\it time} of the peaks in all band passes is close to matching, even if not the same magnitude.  In this case the galactic reddening is 0.024 and additional E(B-V) of 0.4 was added. A constant of 3.0 mag and 2.0 mag were added to the UV and $u$-band light curves. 
The bottom left panel shows the 4 \Msun\, shell removed model with the Type IIb SN2008ax \citep{rom09}. The $uvm2$ observations are only an upper limit indicated by the arrows.  The peak is well fit in the optical and only a few days early in the UV. This model required an additional reddening of 0.4 and UV and $u$-band constants of 2.0 and 1.0 mag respectively. 
The bottom right panel shows the 6 \Msun\, shell removed model with the Type Ib SN 2007Y \citep{jou07,str09}.  The post peak general slope trend matches, but the peak in the model is too early.  This model also required an added reddening of 0.55 and UV and u-band constants of 2.0 and 1.0 mag respectively.

\section{Summary \& Conclusions}

In this paper we have tested the effects of the amount of mass in the hydrogen-rich envelope and the density structure and consequently stellar radius 
on the appearance of SNe light curves and on the evolution of the shock wave as it propagates though the layers of the star.
This test was 
done in two ways: 1) to remove the outer layers of the shell, which also alters the radius, 
and 2) adjust the hydrogen density in the outer shells, lowering the amount of hydrogen, but maintaining the same radius. 
The shell removal and density changes were done just after the launch of the shock, prior to the shock propagation, meaning the outer layers were ignorant of the interior of the star.  The star at this time is 18.9 \Msun, with a 11.9 \Msun\ shell comprised of 6.2 \Msun\ of hydrogen and 5.6 \Msun\ of helium, having lost mass from the initial 23~\Msun.  The amount of bulk mass removed from the shell varied from nothing removed (baseline model) up to 12.12 \Msun.  In the altered density models, 2-10~\Msun\ of material were removed. 

The evolution of the 2 \Msun\ and 4 \Msun\ shell removed models appear very similar to the baseline model for the first month, after which differences begin to develop.  As more and more hydrogen is removed, we would expect a transition from Type IIP's to Type IIb's and eventually  producing Type Ib/c's.  It is interesting that in the shell models there is not a significant change in the first month until 6 \Msun\ is removed (Figure 3), highlighting the need for late time observations.  Models with thicker shells all have similar early time light curves.  As more mass is removed, there is a transition to where the luminosity falls off earlier as more mass is removed, which resembles the Type IIb and Ib/c light curves.  The altered density models follow a similar trend as the shell models, but the post-peak decline is much shallower.  The peak in the light curve in all the models is earliest in the bluest filters as would be expected for thermal cooling.  The most massive star shows the latest light curves peaks, with the least massive star's light curves peaking the earliest. The effect of the mass loss is most pronounced in the UV band passes.    

\begin{figure}[ht]
\center 
\includegraphics[scale=0.40]{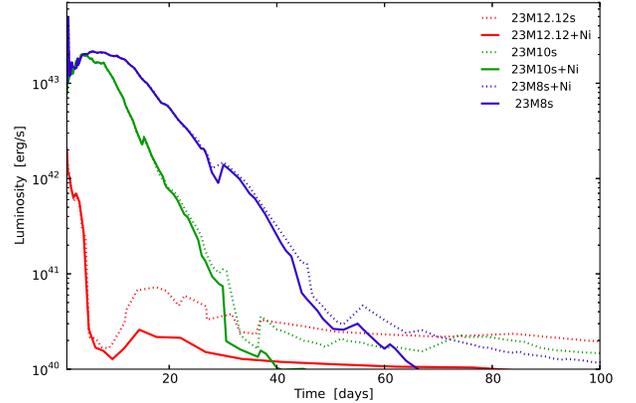}
\caption{Bolometric light curve models with 0.37 \Msun\ of Ni (dotted) compared to the shell removed models of the same mass.}
\label{modelNI}
\end{figure}

It is clear that light curve models covering a large range of peak luminosity and durations can be created with a single progenitor model. The amount and method used to cause mass loss has a significant impact on the light curve.  These models can ultimately be constrained by observations.  The {\it Swift} SNe database \citep{pri13} is an excellent source for optical and UV light curves of more than 50 CCSNe, including complete UV light curves for many CCSNe.  If the removal of hydrogen in the models represents a change from the Type IIP to Type Ib/c, we should see this trend reflected in the data. This general trend is demonstrated in Figure \ref{lcmodplot}, but more modeling is needed to adjust the peak times and the luminosity.  Additionally, we tested 2 models with a higher Ni mass.  These were the same as 23M8s, 23M10s, and 23M12.12s, but with less fall back giving a Ni mass of 0.37~\Msun.  Figure \ref{modelNI} shows the bolometric light curves for these models and the low Ni mass counterparts.  The light curve is identical until about day 10.  Then, as expected, the light curve remains more luminous. The model with 12.12~\Msun\ removed shows a double peak with a sudden brightening of several magnitudes.  Further modeling along this avenue would allow for a better fit to observed SNe light curves in the post-peak decline and to determine the cause of the extreme mass loss double peak. 

With the advent of time-sensitive surveys (e.g. PTF), we are now discovering more rapid transients as well as rapidly varying features in traditional supernovae.  For example, shock breakout, which occurs in the first few days of a supernova explosion has a rise/fall timescale ranging from $<$10 minutes to $\sim$1 day.  These timescales are extremely sensitive to the stellar radius and shock breakout making this time domain an ideal probe of the radial variations caused by mass loss.  Another example arises from the discovery of new, short duration ($<$10 days) transients, which can be modeled with a fully stripped star that lacks a strong wind.  These outbursts are ideally suited to probe extreme mass loss in stars.
 
We have shown from first principles that removing hydrogen from the outer shell does convert a Type IIP to a IIb/Ib. Changing the mass loss does not change the overall luminosity and the models must be fainter and redder to match SN 2010F, SN 2012aw, 2008ax, and 2007Y as well as similar SNe probably because our models were too energetic for these specific SNe.   However, the change in mass loss does show the correct trend to transition from Type IIP to Type Ib/c.  In future modeling, we will need to consider other progenitor stars and parameters, but conclude that the mass loss and method of mass loss are a key component to light curve shape.

\acknowledgments
We thank Joseph Smidt of T-2 at LANL for developing the python scripts used in our post-processed codes and to make plots. Work at LANL was done under the auspices of the National Nuclear Security
Administration of the U.S. Department of Energy at Los Alamos National Laboratory under
Contract No. DE-AC52-06NA25396. All RAGE and SPECTRUM calculations were
performed on Institutional Computing (IC) platforms Wolf, Lobo, and Pinto at LANL.  The work at SwRI was supported by the Internal Research Development (IR\&D) Program, contract \# 15-8333. The authors would also like to thank the anonymous referee for their comments towards improving this paper.

\appendix

\section{Numerical Resolution}
\begin{figure*}
\center 
\includegraphics[scale=0.50]{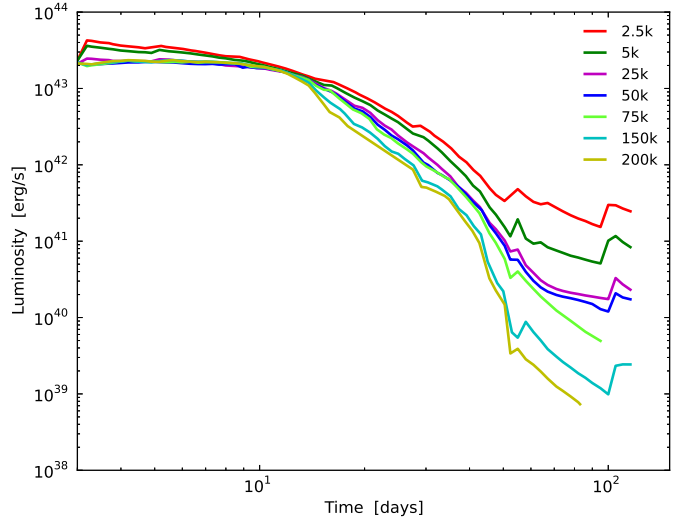}
\caption{Bolometric light curves for the 23M8s model run with several different resolutions in RAGE}
\label{Fapp1}
\end{figure*}

Here the issues related to numerical resolution are investigated in more detail.  The simulation 23M8s is used as a representative model for all simulations presented in this work.   The simulations presented above were all run with a resolution of 100,000 cells at the coarsest resolution and up to 3 additional levels of refinement.  This is significantly more cells than are used in other light curve simulation codes, which are typically Lagrangian.  Recent simulations by other groups include Type 1a SNe  with PHOENIX and SuperNu using 64 zones \citep{supernu}, Pair Instability SNe using Stella with ~250 zones \citep{stella}, and Pair Instability SNe using CMGEN with 150 zones \citep{cmfgen}.   Assuming the Lagrangian cells are distributed uniformly in mass, the mass resolution of these simulations ranges between approximately 0.02 and 1 \Msun.  In comparison to the Lagrangian simulations the mass of cells in 23M8s ranges from 1.0$^{-12}$ to 0.066 \Msun.  Figure \ref{Fapp1} shows the bolometric light curve for the 23M8s model run with resolutions ranging from 2,500 to 250,000 zones.  The simulations all include three levels of refinement in the AMR mesh.  With the exception of the lowest 2 resolutions the simulations are in excellent agreement during the duration of peak brightness.  At 13 days the light curve enters into a rapid decline.  It is at this point that the light curves with greater than 25,000 zones begin to diverge.  By 45 days there is a factor of eight difference in luminosity between the 25,000 and 200,000 zone calculation.  This difference increases to a factor of 50 at 100 days.

\begin{figure*}
\center 
\includegraphics[scale=0.50]{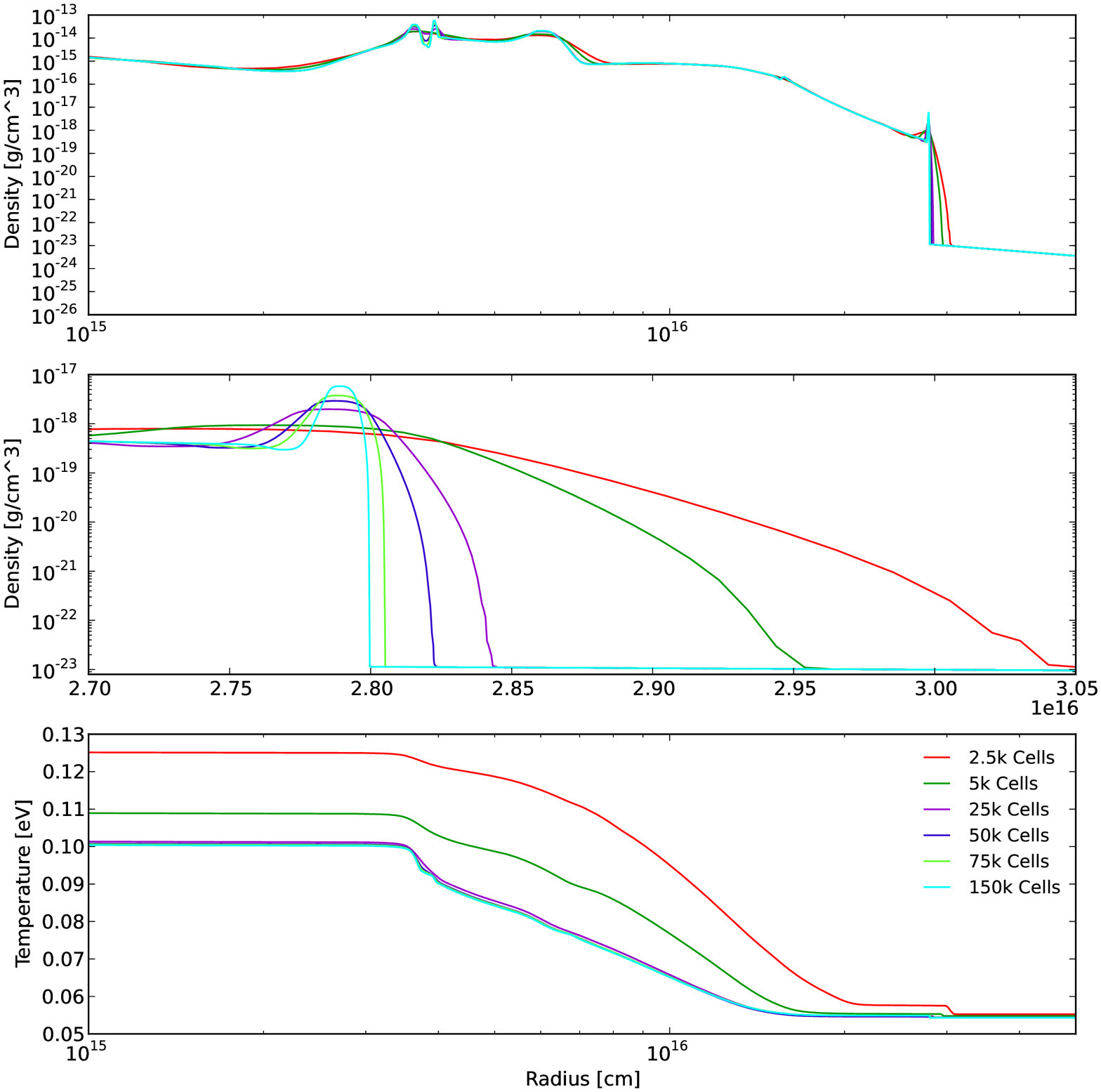}
\caption{The density and temperature of the 23M8s model at 115 days for several different resolutions in RAGE}
\label{Fapp2}
\end{figure*}

Figure \ref{Fapp2} displays the density and temperature for the simulations at 115 days.  Both the density and temperature in the 2,500 and 5,000 zone simulations are significantly different than the other models, which is to be expected from the light curves.  In these two low-resolution simulations several of the key density features are almost completely diffused away and the temperature is 25\% and 10\% higher than in the more converged solutions.  The large scale temperature and density structure is similar for the simulations with greater than 25,000 zones.  The difference in density and temperature between these simulations is less than 1 percent except for near shocks.  The peak density of the leading shock differs by 80 percent between the 75,000 and 150,000 zone simulations.  The steepness of the gradients and peak density in this narrow shock will continue to increase with resolution and if these dominate the region of emission getting a truly converged light curve will be problematic.  However, the cleanness of these shocks itself is an artifact of conducting a 1D simulation.  Figure 15\footnote{See publication in Astrophysical Journal for simulation figure.} displays 2D models of 23M8s model that were run with radiation turned off  in order to make the simulation time tractable.  The simulations were run with 3 different resolutions.  The density is plotted at 115 days.  Similar to the 1D simulations it is clear that the width of the shock is increasing with resolution, however the increase in resolution also leads to instabilities that deform the shape of the shock.  The 2D simulation was started from a perfectly spherical initial model and the instabilities are seeded by numerical effects, in reality the supernovae and star would not be spherical and this would lead to even greater instabilities in the solution.  The 1D simulations inherently prohibit these instabilities from forming and as a result the 1D simulation can generate series of shocks that reverberate through the ejected material and cause sharp effects in the light curve.  It is clear from Figure 15 that in reality the density structure is significantly more complicated and that further refining the 1D simulation will not necessarily create a solution that will be a more accurate match to real supernovae. 

The light curves presented in this paper are post processed separately through the SPECTRUM code.  The SPECTRUM calculation down samples the RAGE grid onto a smaller unstructured mesh.  Figure \ref{Fapp4} shows the light curve for the 23M8s model calculated with three different numbers of zones in SPECTRUM.  The three different resolutions in SPECTRUM produce excellent agreement with bolometric light curves that differ by no more than 10 percent.  The calculations in the main body of the paper were all conducted with 6,000 zones in SPECTRUM. 

 \begin{figure*}[ht]
 \caption{The density structure of two dimension hydrodynamics only simulations for 23M8s at 115 days for 3 different resolutions  }
 \end{figure*}


\begin{figure*}
\center 
\includegraphics[scale=0.50]{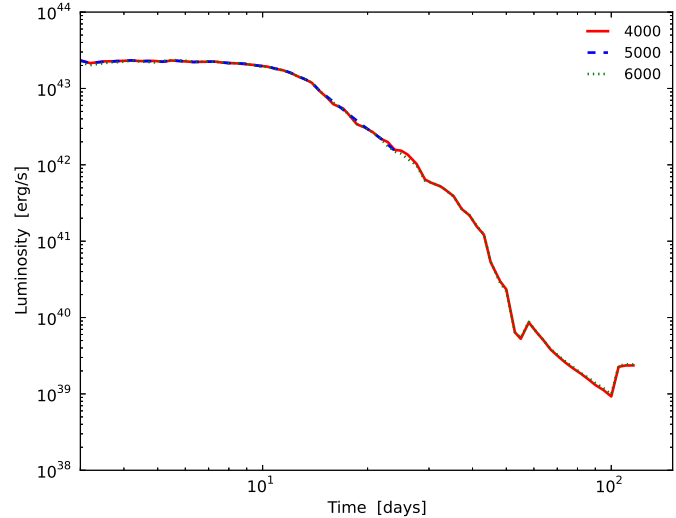}
\caption{Bolometric light curves of 23M8s run with three different resolutions in SPECTRUM}
\label{Fapp4}
\end{figure*}

\newpage
\bibliographystyle{apj}
\bibliography{apj-jour,ragebib}

\end{document}